\documentclass[debug]{rmaa}

\usepackage{paralist}

\usepackage{psfrag,color}

\usepackage[latin1]{inputenc}

\usepackage[bookmarksnumbered=true]{hyperref} 
\hypersetup{
     colorlinks = true,
     linkcolor = blue,
     anchorcolor = blue,
     citecolor = blue,
     filecolor = blue,
     urlcolor = blue
     }

\title{Comprehensive analysis of TESS Full Orbital Phase Curve of WASP-121\MakeLowercase{b}}

\author{
  M.Eftekhar\altaffilmark{1}}

\altaffiltext{1}{Department of Physics, Faculty of Science, University of Zanjan, Iran.}

\shortauthor{Eftekhar, Collaborator, \& Author}
\shorttitle{RevMexAA Main Journal Demo Document}

\fulladdresses{
\item Mohammad. Eftekhar: Department of Physics, Faculty of Science, University of Zanjan, P. O. Box 313 - 45195, Iran
  (m.eftekhar@znu.ac.ir).}

\listofauthors{M. Eftekhar, A. Collaborator, \& L. Author}
\indexauthor{Eftekhar, M.}
\indexauthor{Collaborator, A.}
\indexauthor{Author, L.}

\abstract{We present the full phase curve analysis of the ultrahot Jupiter WASP-121b ($R_p \simeq 1.865 R_J, M_p \simeq 1.184 M_J $) using observations from the Transiting Exoplanet Survey Satellite (TESS) and a comparison between our results with previous studies on this target.
Our comprehensive phase curve model includes primary transit, secondary eclipse, thermal emission, reflection, and ellipsoidal tidal distortion, which are jointly fit to extract the information of all parameters simultaneously from the data sets. We also evaluated and calculated the amplitude of Doppler beaming to be $\sim 2$ ppm, but given the precision of the photometric data, we found it to be insignificant. After removing the instrumental systematic noise, we reliably detect the secondary eclipse with a depth of $489_{-10}^{+16}$ parts-per-million (ppm), dominated by thermal emission. Using the TESS bandpass, we measure the dayside $2941_{-150}^{+61} K$ and nightside $2236_{-97}^{+38} K $ temperatures of WASP-121b. We find that a hotspot is well aligned with the substellar point, leading to the conclusion that there is an inefficient heat distribution from the dayside to the nightside. Our estimated geometric albedo, $A_g = 0.069_{-0.02}^{+0.06}$, suggest that WASP-121b has a low geometric albedo. Finally, our estimated amplitude of the ellipsoidal variation signal is in agreement with the predictions of the theoretical expectations.}

\addkeyword{stars: planetary systems}
\addkeyword{stars: individual: WASP-121}
\addkeyword{techniques: photometric}
\addkeyword{methods: data analysis}

\begin{document}
\maketitle

\section{Introduction}
\label{sec:intro}



Since August 2018, the Transiting Exoplanet Survey Satellite \citep[TESS, ][] {Ricker-tess} has been delivering high-precision photometric observations in a broad optical band (0.6 - 0.95 $\mu m$) for a large sample of bright stars from the southern and northern hemispheres. The wavelength coverage of TESS allows measurements of the combined reflected and thermally emitted planetary light as a function of longitude.  

The exoplanet WASP-121b was discovered by \citet{Delrez2016}, with a period of $\sim 1.275$ days, and it is one of the hottest transiting planets known to date. Its bright host F6-type star (V = 10.4), short orbital period, and inflated radius ($a/R_s = 3.674, R_p = 1.865 R_J$) makes it one of the best targets for investigating its atmosphere with various techniques. Moreover, due to its short orbital period, it is likely that WASP-121b is tidally locked to its host star \citep{daylan}, which makes it probable to have atmospheric features detectable in the averaged planetary flux \citep{ShowmanandGuillot}. 
Several studies have measured WASP-121b's primary transit (when an exoplanet passes in front of its host star) \citep[e.g.,][]{Delrez2016,Evans2016,Evans2018}. By using optical and near-infrared photometry, the depth of its secondary eclipse (i.e., when an exoplanet is occulted by its host star) was determined by \citet{Delrez2016,Kovacs2019,Garhart2019}. The dayside and nightside temperatures of WASP-121b were measured to $2870 K$ and $<2200K$, respectively, according to an analysis of the thermal emission \citep{Bourrier}, which is to be expected given the planet's proximity to its host F-type star. The reflection component was not included in \citet{Bourrier} phase curve, but we take it into account in our comprehensive full phase curve model. WASP-121b's geometric albedo was estimated as $0.070_{-0.040}^{+0.037}$ based on the optical phase curve analysis by \citet{daylan}. In our analysis, the reflection component is also calculated simultaneously with other parameters to highlight the correlations between all of the constrained parameters.

The main objective of the current study is to learn more about the thermal emission and atmospheric structure of WASP-121b by performing our comprehensive joint model and comparison our results with
previous studies like \citet{daylan, Bourrier}.
To achieve this, we analyze the full-orbit optical TESS phase curve and model the primary transit, secondary eclipse, and four main components of the phase curve, which include the tidal ellipsoidal distortion, thermal, and reflected emission of the planet. We also calculated rotational Doppler beaming and discovered that it isn't significant given the precision of the light curves. We can determine the uncertainty and correlations among all constrained parameters using our comprehensive joint model, which allows us to extract information from all parameters at the same time.


Here, we describe our WASP-121b analysis by presenting our comprehensive phase curve model and comparing our findings to previous measurements.
The paper is organized as follows; in Section \ref{sec:Observations} we describe the observations and data reduction methods that were used. In Section \ref{sec:Phase curve} we describe in detail the four different components that were used to characterise the phase curve of WASP-121b. In Section \ref{sec:model} we present our joint model as well as the fitting procedure we employed to acquire our results. We provide our physical parameters derived from TESS observations in Section \ref{sec:result}, and discuss with a brief summary in Section \ref{sec:Discussion}.

\section{Observations and data reduction}
\label{sec:Observations}

Between the 8th of January and 1st of February 2019, the TESS camera 3 monitored WASP-121 (also known as TIC 22529346) throughout its sector number 7. The observation span was 24.5 days and included 18 primary transits of WASP-121b. 

Photometric data were processed through the Science Processing Operations Center (SPOC) pipeline \citep{Jenkins2017}. In this study, we decided to use PDCSAP (Pre-Search Data Conditioning) light curves because they are corrected for instrumental systematic noise which is present  in the Simple Aperture Photometry (SAP) light curves; thus PDCSAP light curve show considerably less scatter and short-timescale flux variation \citep{Smith2012,Stumpe2014}. PDCSAP light curve of WASP-121 was also used in other studies investigating the phase curve of WASP-121b such as \citet{Bourrier}.

\begin{figure}[!t]
  \includegraphics[width=\columnwidth,  height=9 cm]{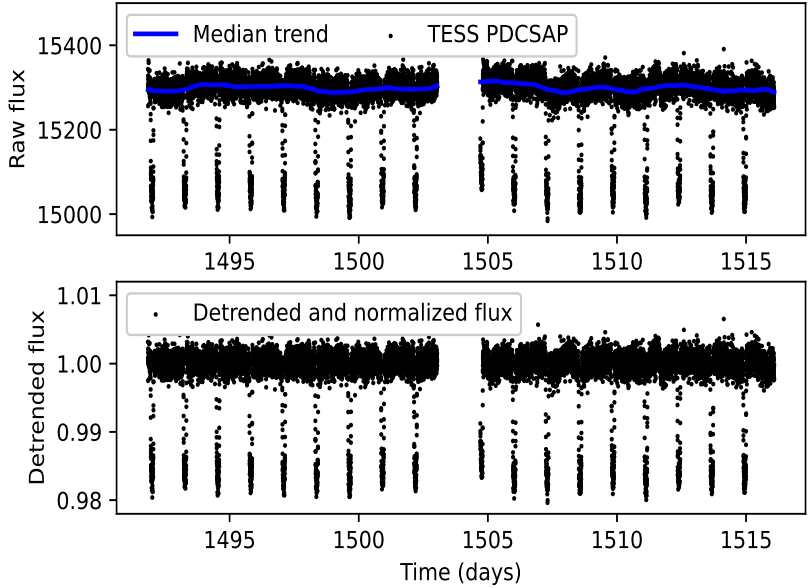}
  \caption{(Top) TESS light curve (PDCSAP flux) of  WASP-121. The PDCSAP photometry is indicated with black dots, and the solid blue line shows the trend obtained by applying a detrending filter determined by \texttt{wotan}. (Bottom) PDCSAP light curve after normalization by its median detrending.}
  \label{fig:deterended}
\end{figure}

The PDCSAP photometry is presented in the upper panel of Figure \ref{fig:deterended}, which shows the remaining systematics in the data at short time scales, particularly in Sector 7's second orbit. Instrumental effects including changes in the thermal state of TESS and pointing instabilities cause these remaining systematics.
The PDCSAP light curve's median was used to normalise the data. 
To have a fair comparison with \citet{Bourrier}, we did exactly the same steps in preprocessing of data. Although the dominant systematics were corrected by default in the PDCSAP light curve, we corrected it further for the remaining systematics. To do this, we used the median detrending algorithm with a window length of one orbital period to smooth the PDCSAP light curve, keeping variability at the planetary period and minimising the effect of normalisation on the phase curve of WASP-121b. If we choose a smaller window length of one orbital period, then it is very likely that the signal will be absorbed and removed from the atmosphere. We followed the same processes using \citet{Bourrier}, the regression is shown in Figure \ref{fig:deterended} and was implemented using the Python package \texttt{wotan} \citep{Hippke}. We also performed phase folding at the orbital period of WASP-12b and binned every 50 datapoints after detrending; this reprocessed data was used in our further analysis. The reprocessed data is shown in Figure \ref{fig:fold}.

\section{Phase curve}
\label{sec:Phase curve}
In addition to the primary transit light curve and secondary eclipse, photometric observations reveal additional variation induced by the orbiting exoplanet over the full planetary orbit. This variation can be decomposed into several components, namely thermal emission, reflected light, Doppler beaming, and ellipsoidal variation. In this study, we assume that the phase curve variation is a combination of thermal emission, reflection, and ellipsoidal variation, and we ignore the Doppler beaming, which the reason will be explained in Subsection \ref{sec:Doppler}. 

\subsection{Thermal emission}
\label{sec:Thermal emission}
As mentioned in Section \ref{sec:intro}, due to its tidal locking and proximity to its host star, inefficient heat transport from the dayside to the nightside WASP-121b should have a significant temperature difference between its permanent day and night sides \citep{Bourrier}. As a result, WASP-121b is expected to have a zone (hotspot) with maximum temperature and higher thermal flux in comparison to the rest.

In order to model WASP-121b's thermal emission component, we used a semi-physical model based on \citet{Zhang2017} which has been implemented in \texttt{spiderman} \citep{luden2018}. 
It uses three parameters to reproduce the main characteristics of the thermal light curve. The thermal phase shift is controlled by the ratio of radiative versus advection time scale, $\xi$. The hotspot's longitudinal shift becomes larger as $\xi$ increases. If $\xi$ increases, the nightside temperature increases while the dayside temperatures drop, resulting in a reduction in the difference between day and nightside temperatures. The temperature on the planet's night side is controlled by the nightside temperature, $T_{N}$. Finally, $\Delta T_ {DN}$ represents the difference between day and night temperatures.
To calculate the temperatures in the TESS bandpass, we used Phoenix model spectrum \citep{Husser2013} for the host star by using \texttt{spiderman}.


\subsection{Reflection}
\label{sec:Reflection}
In the bandpass of observations, the reflection is the proportion of light from the host star that is reflected by the planetary atmosphere and/or planetary surface. The phase modulation of the reflection is sinusoidal, with the same maximum and minimum as the thermal emission. The difference in reflectivity (albedo) determines the amplitude of the reflection \citep{Shporer2017}. A basic form of reflection phase modulation can be described as:
\begin{equation}
  \label{eq:one}
Reflection =  A_{ref}(1+\cos(2\pi (\phi +\Delta_P /P)+\pi)),
\end{equation}
where, $A_{ref}$ is the amplitude of the reflection, which depends on the albedo, $\phi$, is the orbital phase, $P$ is the orbital period, and $\Delta_P$ is the phase shift. 
The geometric albedo of a planet, $A g$, is the ratio of its reflectivity at zero phase angle to that of a Lambertian disk, and can be calculated as \citep{Rodler2010}
\begin{equation}
  \label{eq:two}
A_g = A_{ref}(a/R_p)^2,
\end{equation}
where $a$ is the semi-major axis and $R_p$ is the planet's radius.

\subsection{Doppler Beaming}
\label{sec:Doppler}
Doppler Beaming is caused by relativistic effects on the host star's emitted light along our line of sight. For circular orbits, the Doppler beaming component has a sinusoidal form with a maximum during the quadrature ($0.25$) phases and at the quadrature ($0.75$) phases.
The amplitude of the beaming component, $A_{beam}$, can be computed using the physical parameters of the system as \citet{Shporer2017}
\begin{equation}
  \label{eq:three}
A_{beam} = 0.0028\alpha_{beam}(\frac{P}{day})^{-1/3}\times (\frac{M_1+M_2}{M_\odot})^{-2/3}(\frac{M_2\sin i}{M_\odot}),
\end{equation}
where
\begin{equation}
  \label{eq:four}
\alpha _{beam} = \int \frac{1}{4}\frac{xe^{x}}{e^{x}-1}dx , x\equiv \frac{hc}{kT_{eff}\lambda }.
\end{equation}
Here $M_1$, $M_2$, $M_\odot$ are the masses of the host star, planet, and sun, respectively. $i$ is the orbital inclination angle, $h$ is Planck's constant, $k$ is Boltzmann's constant, $T_{eff}$ is the stellar effective temperature, and $\lambda$ is the observed wavelength.

In our study, this integral should be taken in the TESS passband. Based on Equation \ref{eq:two}, we estimate the amplitude of Doppler beaming to be $\sim 2$ parts-per-million (ppm), which is significantly smaller than the precision achievable by TESS (even for the case of a star as bright as WASP-121), so we decided to exclude the Doppler beaming from our total phase curve model.

\subsection{Ellipsoidal variations}
\label{sec:ellipsoidal}
The gravitational pull of a close-in exoplanet causes the host star to deviate from a spherical form to an ellipsoid. This deformation produces photometric orbital modulations with an amplitude that can be approximated by \citet{Shporer2017}

\begin{equation}
  \label{eq:five}
A_{ellip}\simeq 13\alpha _{ellip}\sin i\times (\frac{R_1}{R_\odot})^{3}(\frac{M_1}{M_\odot})^{-2}(\frac{P}{day})^{-2}(\frac{M_2\sin i}{M_J})[ppm],
\end{equation}
where 
\begin{equation}
  \label{eq:six}
\alpha _{ellip} = 0.15\frac{(15+u)(1+g)}{(3-u)}.
\end{equation}
Here, $u$ is the linear limb darkening coefficient and $g$ is the gravity darkening coefficient. We utilized a tabulation of these coefficient values from \citet{claret2017A&A} and estimated the amplitude of ellipsoidal variation to be $\sim 20$ ppm, which is compatible with the precision achievable by TESS on WASP-121. Therefore, we decided to consider the ellipsoidal modulations in our total phase curve model. The ellipsoidal variation shows two peaks at phase quadratures $0.25$ and $0.75$, respectively, and can be modeled as:
\begin{equation}
  \label{eq:seven}
Ellipsoidial = A_{ellip}(1+\cos(4\pi\phi-\pi))
\end{equation}

\section{Model and Fitting procedure}
\label{sec:model}

Our joint model consists of primary transit, secondary eclipse, and phase curves that incorporate the thermal emission, reflection, and ellipsoidal variations. We also included a constant baseline to compensate for any normalization bias. For the primary transit and secondary eclipse we used the Python packages \texttt{batman} \citep{batman} and for the thermal emission, we used \texttt{spiderman} \citep{luden2018}. Our thermal model is based on a semi-physical model of Zhang \& Showman (see Subsection \ref{sec:Thermal emission}) implemented by \texttt{spiderman}. The reflection is modeled as Equation \ref{eq:one} (see Subsection \ref{sec:Reflection}) and ellipsoidal variations is modeled as Equation \ref{eq:seven} (see Subsection \ref{sec:ellipsoidal}). Performing a joint model analysis allows us to extract information about all parameters simultaneously from the data sets. It also gives us the ability to assess the uncertainty and correlations between all of the constrained parameters.

To determine the parameters of the full phase curve, we fitted our joint model to the reprocessed data (see Section \ref{sec:Observations}). The best fit parameters and their associated uncertainties are determined using a Markov Chain Monte Carlo (MCMC) approach using the affine invariant ensemble sampler \texttt{emcee} package \citep{Foreman-Mackey2013PASP}.

We fit for $R_p/R_s$, $a/R_s$, $i$, $u_1$, $u_2$, $\xi$, $T_N$, $\Delta T_{DN}$, additive baseline, secondary eclipse depth, $A_{ref}$, $\Delta_P$ and $A_{ellip}$. The priors of $u_1$, $u_2$, $\xi$, $T_N$, and $\Delta T_{DN}$ are equal to \citet{Bourrier}. The priors of $R_p/R_s$, $a/R_s$, and $i$ had normal priors in \citet{Bourrier}, and we chose an uninformative uniform prior for them. The additional parameters in our model have a wide uninformative prior, allowing us to obtain their best estimation.

Table \ref{tab:tab-one} provides information on individual prior distributions that were chosen. We fix the transit epoch, $T_0$ and orbital period, $P$ because we use one sector of TESS data which covers about 24 days, whereas the period from \citet{Delrez2016} takes into account years of WASP data, which provides more information on the period. Considering the Lucy-Sweeney bias \citep{Lucy&Sweeney}, we adopt a circular orbit by fixing the eccentricity $e$, to 0, and the argument of periastron, $\omega$ to values obtained by \citet{Bourrier}. To generate the posterior distributions, we ran 700 walkers over 1000 steps with a burn-in phase of the $20\%$ sample. The walkers are plotted and visually inspected for convergence. We estimated the median and standard deviation from the posterior distributions at $1\sigma$, which contains $68\%$ of the posterior distribution, for our best fitted values and uncertainties.

\begin{table}[!t]\centering
  \setlength{\tabnotewidth}{0.5\columnwidth}
  \tablecols{3}
  \setlength{\tabcolsep}{2.8\tabcolsep}
  \caption{Free parameters, uniform priors range, and the best fitted values.} \label{tab:tab-one}
 \begin{tabular}{lrr}
    \toprule
    Parameter & \multicolumn{1}{c}{Prior} & \multicolumn{1}{c}{Value} \\
    \midrule
    Planet-star radii ratio; $R_p/R_s$ & $[0, 1]$  & $0.1234_{-0.0005}^{+0.0005}$\\
    Scaled semi-major axis; $a/R_s$  & $[0, 5]$ & $3.792_{-0.039}^{+0.023}$\\
    Orbital inclination $i$ (deg) & $[0, 90]$ & $88.80_{-1.23}^{+1.27}$\\
    limb darkening coefficient; $u_1$ & $[0, 1]$ & $0.260_{-0.042}^{+0.034}$\\
    limb darkening coefficient; $u_2$ & $[0, 1]$ & $0.132_{-0.082}^{+0.056}$\\
    Radiative to advective timescales ratio; $\xi$  & $[-10, 10]$ & $-0.022_{-0.141}^{+0.159}$\\
    Nightside temperature; $T_N$ (K)  & $[0, 5000]$ & $2236_{-38}^{+97}$\\
    Day-night temperature difference; $\Delta T_{DN}$ (K) & $[0, 2000]$ & $734_{-55}^{+28}$\\
    Additive baseline & $[-0.1, 0.1]$ & $-0.00014_{-0.8 \times 10^{-6}}^{+1.7\times 10^{-6}}$\\
    Secondary eclipse depth (ppm) & $[0, 800]$ & $489_{-10}^{+16}$\\
    Amplitude of the reflection; $A_{ref}$ (ppm)  & $[0, 500]$ & $73_{-3.1}^{+2.2}$\\
    Reflection phase shift; $\Delta_P$ & $[-0.5, 0.5]$ & $-0.0008_{-0.0071}^{+0.0012}$\\
    Amplitude of ellipsoidal variations; $A_{ellip}$  (ppm) & $[0, 100]$ & $20_{-3}^{+2}$\\
    \bottomrule
  \end{tabular}
\end{table}

\begin{figure}[!t]
  \includegraphics[width=\columnwidth, height=11 cm]{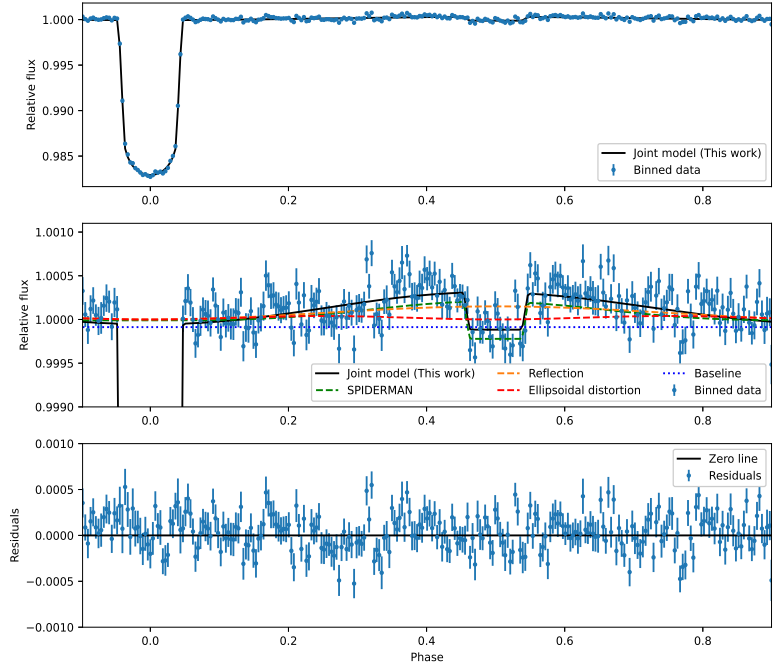}
  \caption{(Top) Our reprocessed data of WASP-121 (blue dots) compared to our best fitted full phase curve model (black curve). (Middle) zoom of the secondary eclipse and phase curve variations with the reflection modulation (dashed orange curve), ellipsoidal distortion (dashed red curve), and baseline (dotted blue line). (Bottom) The best fitted model's corresponding residuals.}
  \label{fig:fold}
\end{figure}

\begin{figure}[!t]
  \includegraphics[width=\columnwidth]{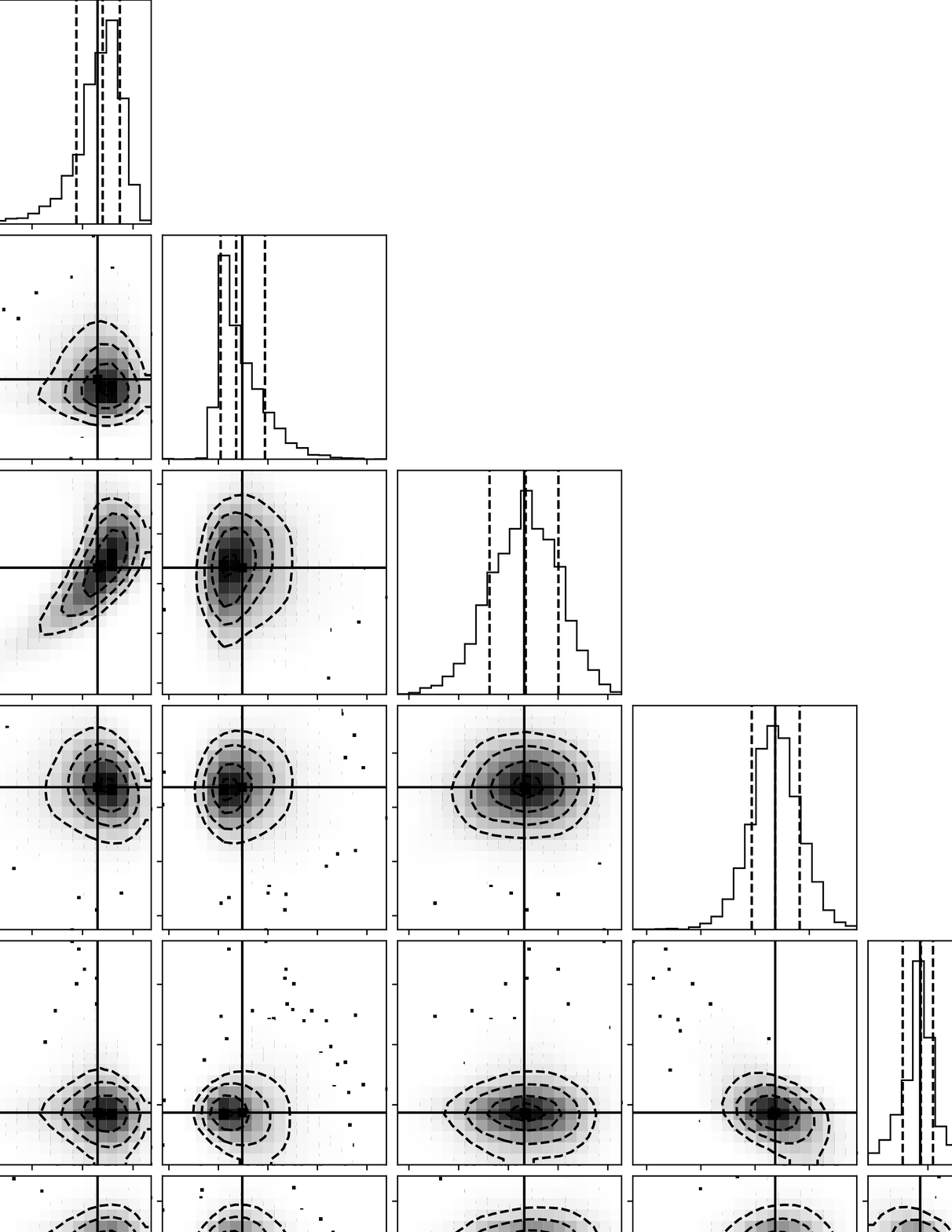}
  \caption{Retrieved posterior distributions by fitting our joint model to the phase curve of the WASP-121b. the black points indicate the best-fit values, and the colours of the contours
  highlight the $1$ and $2 \sigma$ simultaneous $2D$ confidence regions which contain respectively $39.3\%$ and $86.5\%$ of the samples. The solid black line corresponds to the median values, while dashed black
  lines show the $1 \sigma$ highest density intervals.}
  \label{fig:corner}
\end{figure}

\section{result}
\label{sec:result}

The results of our joint model fitting to the reprocessed data are shown in Table \ref{tab:tab-one}. The best fitted model's reduced chi-squared $\chi2$ (i.e., RMS of the residuals per degree of freedom) is $1.29$, indicating a good fit to the TESS photometry. Figure \ref{fig:fold} shows the reprocessed data, as well as the best fitted model of the full phase curve. Our best-fitted model's residuals still exhibit some correlated noise, which could be uncorrected TESS systematic noise. Identical correlated noise was also present in the residuals of the best-fitted \citet{Bourrier}. The corner plot for our retrieved posterior distributions from the joint model fit is shown in Figure \ref{fig:corner}.
In addition, when we fit our joint model to unbinned data, we found that our results are generally consistent. Furthermore, we binned every 80 datapoints and fitted our joint model, and the results were consistent with the values presented in table \ref{tab:tab-one}. This test demonstrates that the results provided in this work are robust to the binning effect.

We calculated a planetary radius (in stellar radii), $(R_p/R_s)$, of $0.1234_{-0.0005}^{+0.0005}$ and a reasonably large secondary eclipse depth with amplitudes of $489_{-10}^{+16}$ ppm. Our measured secondary eclipse depth value is within $1 \sigma$ of the value reported in the \citet{daylan}. However, our estimated value is larger ($1.6 \sigma$) than \citet{Bourrier} measured value. Other orbital and transit parameters agree well with previously published values in the literature \citep{Wong, daylan, Bourrier}.

Our estimation on geometric albedo using Equation \ref{eq:two} is $0.069_{-0.02}^{+0.06}$ which is consistent with the estimate of  \citet{daylan} of $0.070_{-0.040}^{+0.037}$ ppm. \citet{Mallonn2019} estimated geometric albedo of $A_g = 0.16 \pm 0.11$ in the $z^\prime$ band.

We measured the ratio of radiative versus advection time scale of atmospheric height is $\xi = -0.022_{-0.141}^{+0.159} $, which is consistent with zero. This implies that there is no thermal redistribution between WASP-121b's day and night sides, resulting in a larger day-night temperature difference. We measured the temperature of the night and day sides, $2236_{-38}^{+97} K $ and $2941_{-61}^{+147} K$, respectively, which are in agreement with the values published in \citet{Bourrier} and \citet{daylan}. \citet{Parmentier2018} and \citet{Evan2017Natur} by fitting the blackbody model to Spitzer and Hubble Space Telescope WFC3 observations could measure the dayside temperature of $2650 \pm 10 K$ and $2700 \pm 10 K$, respectively.

Based on our best fitted model, we estimated the ellipsoidal variation amplitude to be $20_{-3}^{+2}$ ppm, which is more in line with the theoretical estimate of $20$ ppm based on Equation \ref{eq:five} and slightly larger than \citet{daylan} estimation which was $8_{-6}^{+12}$ ppm.

The most remarkable result of our study is the simultaneous measurement of the primary transit, the secondary eclipse, and the robust detection of the total phase curve component corresponding to thermal emission, reflected light, and ellipsoidal variation (see table \ref{tab:tab-one}). The three components and full-phase curve are plotted in the middle panel of Figure \ref{fig:fold}.

In our analysis, we also experimented what would happend if we let the eccentric $e$ and $\omega$ free in our joint fit. In this case we obtained that these results are consistent with the values reported in Table \ref{tab:tab-one} at about $\sim 1 \sigma$. We obtain eccentricity constraints: $e = 0.0024_{-0.0024}^{+0.0041}$ and $\omega = 9.05_{-1.06}^{+2.32}$ deg which are consistent with those published in \citet{Bourrier}. According to Lucy-Sweeny bias \citep{Lucy&Sweeney}, in order to measure a non-zero eccentricity with $95\%$ confidence, a result of $e > 2.45 \sigma_{e}$ is required, where $\sigma_{e}$ is the standard deviation of the eccentricities \citep{Eastman}. As a result, we can confidently rule out WASP-121b non-zero eccentricity.

In addition to our total phase curve model, we investigated a scenario in which the planetary flux is purely reflective. To approximate the planetary reflection, we used the Lambertian reflection model implemented in \texttt{spiderman} and characterized by a geometric albedo $A_g$. Using this scenario we estimated the geometric albedo to be $0.46_{-0.035}^{+0.036}$ which is significantly ($3 \sigma$) larger than the estimate of the geometric albedo reported by \citet{Bourrier}. It is quite close to $A_g = \delta(a_p /R_p )^{2} = 0.47_{-0.03}^{+0.03} $ which is the value derived from the TESS secondary eclipse depth ($\delta = 489_{-10}^{+16}$ppm). The reduced $\chi^2$ of this purely reflective scenario is $1.8$.

\section{Summary and conclusions}
\label{sec:Discussion}

In this work, we presented our full phase curve model for analyzing the transiting ultra-hot Jupiter WASP-121b utilizing one sector of TESS observations. There were two reasons for using only one sector of TESS. The first is that different TESS sectors have different systematic noises, and combining several sectors may introduce additional complications and difficulties in our joint modelling. The second and most important reason is that we wanted to use the same data set as in \citet{Bourrier} and \citet{daylan} so that we could assess how much improvement we could get from using a more complete model. We first used the median detrending technique with a window length of one orbital period of WASP-121b to conduct smooth detrending on the TESS data, in order to have comparable data with \citet{Bourrier} also did exact same steps.
We binned every 50 data points after phase folding at the orbital period, as \citet{Bourrier} performed previously. In our subsequent analysis, we used this reprocessed data. Then we fitted our joint model to the reprocessed data. Our joint model consists of primary transit, secondary eclipse, and phase curves that incorporate the thermal emission, reflection, and ellipsoidal variations.

We reliably measured the secondary eclipse with a depth of $489_{-10}^{+16}$ ppm after eliminating systematic noise. The combination of thermal emission and reflection in the TESS bandpass results in a relatively significant secondary eclipse depth of WASP-121b. Due to the strong stellar irradiation and low geometric albedo, the secondary eclipse is expected to be mostly dominated by the planet thermal emission.

Our measurement of the $\xi = -0.022_{-0.141}^{+0.159}$ is statistically consistent with zero. This value indicates that the atmosphere of WASP-121b has inefficient thermal redistribution from dayside to nightside, which is consistent with results in the literature \citep{Bourrier, daylan} and with theoretical models \citep{Komacek2017ApJ,Perez-Becker2013ApJ}. 
WASP-121b's maximum temperature region is located near the sub-stellar point (close to the sub-stellar point) due to inefficient thermal redistribution, as advection does not redistribute heat across longitudes \citep{Zhang2017}. The inefficient thermal redistribution also results in substantial differences in the night and dayside temperatures ($734_{-55}^{+28} K$) of WASP-121b. Our measured dayside temperature of $2941_{-150}^{+61} K$ for WASP-121b places it in the ultra-hot Jupiter class \citep{Parmentier2018, Bell2018ApJ}.

In this study, we did not assume that the flux of WASP-121b measured by TESS was exclusively thermal emission, and we took into account reflected light. Our best fitted joint model yielded a low geometric albedo of $0.069_{-0.02}^{+0.06}$, indicating that reflection in the TESS passband of WASP-121b is not negligible, which was ignored by \citet{Bourrier}.
Our estimated low geometric albedo value is consistent with \citet{daylan} and other hot Jupiters, in particular, irradiated hot Jupiters at the same wavelength as \citet{Schwartz2015}. It is also consistent with other short-period hot Jupiter planets, such as WASP-18b ($A_g< 0.048$ at $2\sigma$; \citet{shporer2019AJ....157..178S}), Qatar-2b ( $A_g< 0.06$ at $2\sigma$; \citet{dai2017AJ}), and WASP-12b ($A_g< 0.064$ at $97.5\%$ confidence; \citet{Bell2017}). Considering the fact that the bandpass of TESS is close to the wavelength region where the host star is brightest, the bond albedo is small when the geometric albedo is small \citep{shporer2019AJ....157..178S}.

The amplitude of the ellipsoidal variation and Doppler beaming are significantly smaller than that of reflected light and thermal emission, according to theoretical estimates (see Figure \ref{fig:fold}). We didn't incorporate Doppler beaming in our phase curve model because our theoretical estimation of the amplitude of Doppler beaming yields a value of $\sim 2 $ ppm , which isn't significant given the precision of photometric data.
Finally, our best fitted joint model also provided us with an estimate of the amplitude of ellipsoidal variation that is consistent with theoretical expectations. 

Hot host star and short orbital period of WASP-121b make it to be highly irradiated. Furthermore, the lack of statistically significant phase shift, poor heat distribution, and low albedo are all compatible with other highly irradiated gas giant planets. This study demonstrated that our model may be used to explore the full phase curves of transiting systems. The fact that the WASP-121b phase curve modulations were clearly detected shows that TESS data are sensitive to photometric variations in systems with short periods and massive planets.

More TESS data from extended missions or from other existing facilities like the CHaracterising ExOPlanet Satellite (CHEOPS) \citep{Benz2021} will also enable for a more in-depth study of exoplanets' full phase curve. Our WASP-121b retrieval analysis provides a glimpse into the comprehensive analysis of the full orbital phase curve which can be performed by combining optical and thermal infrared observations near-infrared emission using existing facilities like the ARIEL \citep{Tinetti2018ariel} and upcoming facilities with higher resolution such as the James Webb Space Telescope (JWST) \citep{Gardner2006SSRv..123..485G}.


\section{acknowledgements}
NASA's Science Mission Directorate funding the TESS mission. Our work is based on data collected by this mission, available at Mikulski Archive for Space Telescopes (MAST). Special thanks to Mahmoudreza Oshagh, who helped with useful suggestions that greatly improved the paper, and fruitful discussions on the topics covered in this paper. I would like to thank the referee for very useful suggestions that greatly improved the paper.

\bibliography{Reference3}
\end{document}